# Enhancing Financial Literacy and Management through Goal-Directed Design and Gamification in Personal Finance Application

Phuong Lien To

School of Arts and Creative Technologies, University of York
phuonglionto161@gmail.com

**Abstract.** This study explores the development of a financial management application for young people using Alan Cooper's Goal-Directed Design method. Through interviews, surveys, and usability testing, the Mabu app was designed to improve financial literacy by combining personalised features and gamification. Findings highlight the effectiveness of gamified learning and tailored experiences in encouraging better financial behavior among young users.

## 1 Introduction

As the economic environment is changing at breakneck speed, personal financial management has become more critical than ever [1]. The lack of personal financial management skills increasingly [2] puts young people under economic pressure, unable to manage problems that arise in life, leading to being caught in a spiral of debt [3, 4]. According to the 2023 UK Money Literacy Test, only 17% of individuals aged 18-24 demonstrated sufficient financial literacy, highlighting an urgent need for accessible, engaging financial education.

With mobile applications increasingly used to assist in money management, there is an opportunity to design digital tools that not only track finances but also educate and motivate young users. Gamification - embedding game elements like points, badges, and feedback, has proven effective in driving user engagement across many domains, yet its application in personal finance remains underexplored. In parallel, Goal-Directed Design [5] offers a structured method to align product functionality with users' real goals.

This paper presents the development and evaluation of Mabu, a mobile financial management app designed for young adults. By combining Goal-Directed Design and gamification, Mabu aims to increase financial awareness, user motivation, and promote sustainable financial behaviours through an engaging and user-centric experience.

To guide this work, the following research questions were addressed:

- **RQ1**. How do young people currently manage their finances?
- **RQ2**. What challenges do they face in managing their finances effectively?
- **RQ3**. To what extent can Goal-Directed Design and Gamification increase user engagement and motivation in personal finance apps to improve financial literacy?

## 2 Background and Related Works

### 2.1 Financial Literacy

Financial literacy is a foundational element for achieving financial inclusion and long-term well-being. It influences individuals' capacity to engage with financial products and services [6]. The Organisation for Economic Co-operation and Development (OECD) defines financial literacy as the ability to understand and apply financial knowledge to make informed decisions [1]. Numerous studies have linked low financial literacy to problematic financial behaviors, including high debt levels, limited savings, and poor responses to unexpected expenses [7-9]. Conversely, individuals with higher financial literacy demonstrate better investment decisions, savings behaviors, and retirement planning [10, 11].

Despite its significance, financial literacy levels remain relatively low across both developed and developing contexts [1]. Educational programs targeting youth often struggle with incomplete curricula and lack of learner engagement, resulting in limited real-world applicability [12]. This is especially problematic as young adults frequently experience challenges in money management, which can affect academic performance and overall well-being [13-15]. Recent evidence suggests that digital financial tools can positively impact financial behavior among young users [16], highlighting opportunities to leverage technology to address these gaps.

### 2.2 User Experience and Gamification in Mobile Financial Applications

User experience (UX) design plays a critical role in digital product development. The Technology Acceptance Model (TAM) (Fig. 1) introduced by Davis [17] identifies Perceived Usefulness and Ease of Use as key drivers of user acceptance.

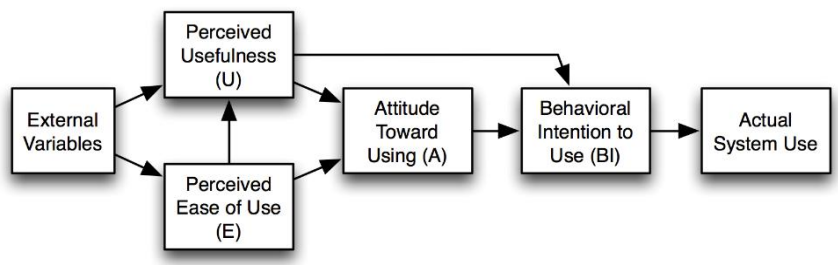

**Fig. 1.** The Technology Acceptance Model [17].

Building on this, the UTAUT model [18] expands the framework with additional constructs: Performance Expectancy, Effort Expectancy, Social Influence, and Facilitating Conditions, along with moderating factors such as age and experience (Fig. 2). Later, UTAUT2 [19] further incorporates Hedonic Motivation, Price Value, and Habit, acknowledging the growing complexity of technology adoption in everyday life.

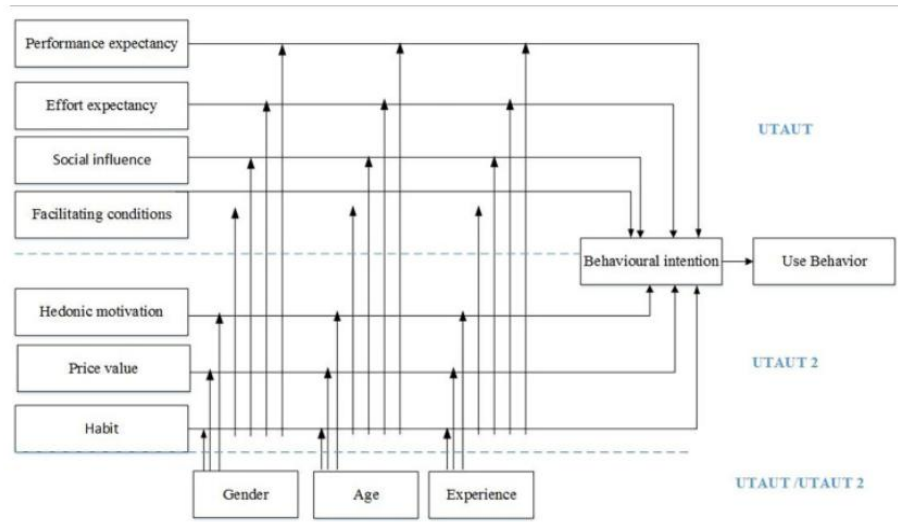

**Fig. 2.** UTAUT and UTAUT2 model [20].

Mobile devices are widely used in human life today. Compared with user experience in other platforms, mobile user experience has differences [21]. Applying technology to raise awareness about personal financial management is a popular trend today with the advent of many software and applications supporting financial management. These apps are like personal money managers for young people as they give users an overview of how their money is spent so that users can monitor and plan spending appropriately [22]. However, these applications only provide tools, and how to use them effectively depends on the user. Therefore, they are not optimised to help users gain financial knowledge to help them proactively build a strategy to manage their accounts.

A key construct in UTAUT2, hedonic motivation, has received growing attention for its ability to enhance user engagement. In the model, hedonic motivation comes from users feeling joy or being rewarded [19]. Espinosa-Curiel et al. [23] have proven the positive influence of children's excitement in acquiring knowledge through playing video games. Similarly, according to the survey results of Hussain et al. [24], enjoyment is one of three factors that impact user adoption of mobile maps.

A popular method of increasing hedonic motivation is using gamification. Hamari et al. [25] define gamification as the use of game-related elements such as points, badges, leaderboards, and rewards in non-game contexts, including mobile application design. This method is used by designers to increase user motivation and interaction with the product or system [26] when it can bring joy and excitement to the user after the targeted task completion.

Gamification is one of the emerging techniques applied to FinTech to help increase user engagement or intention rather than financial benefits [27]. Working together on the application of gamification in the financial sector, Bayuk and Altobello [28] demonstrate the effectiveness of this kind of technology in improving the financial literacy of college students. Financial service applications generally have limited design elements to increase because they are designed to monitor accounts and perform transactions. However, if elements related to playfulness are added, customers will be more interested in using it [29].

### 2.3   Goal-Directed Design Framework

In user experience design, Goal-Directed Design is one of the popular methods of designing products, including digital products. It focuses on meeting the goals of users [30]. Goal-Directed Design was first introduced by Alan Cooper in his book in 1995 [31].

In general, the process of Goal-Directed Design includes six stages: Research, Modeling, Requirements, Framework, Refinement, and Validation [32]. Through user research and persona development, designers gain deep insights into user motivations and contexts. These are then translated into functional and emotional requirements, visualized through frameworks and prototypes, and refined via usability testing before final deployment (Fig. 3).

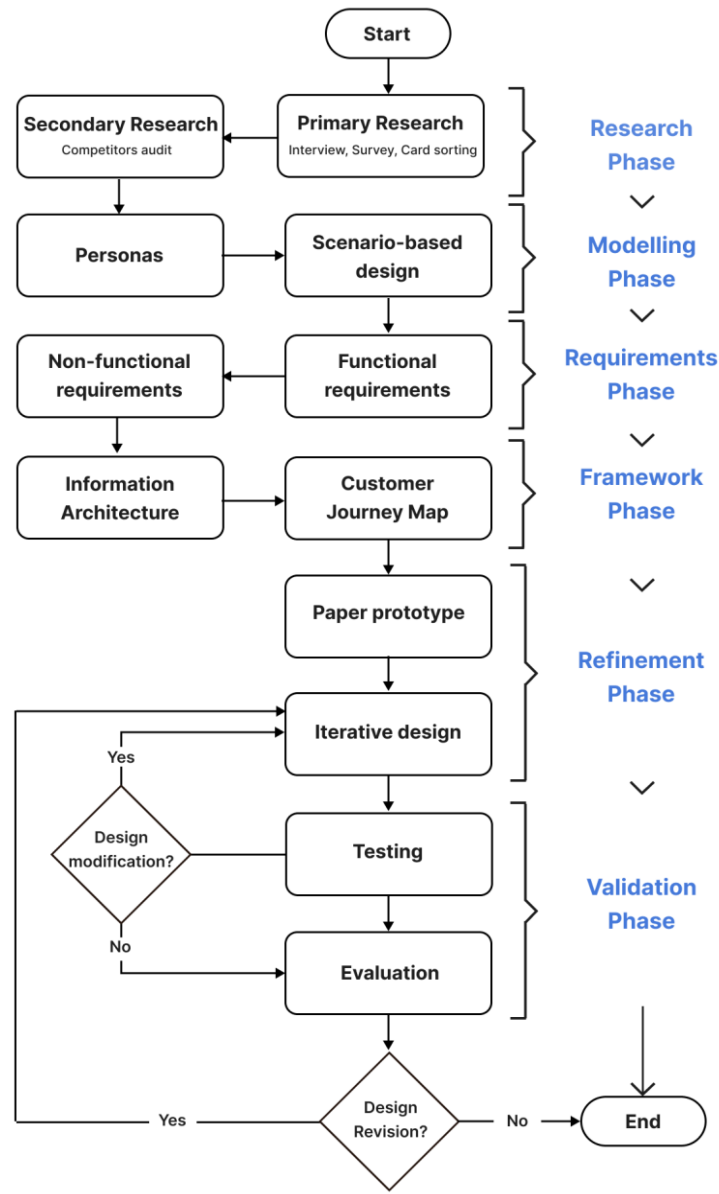

**Fig. 3.** Goal-Directed Design Framework.

Goal-Directed Design is particularly valuable in domains where understanding user intent is critical, such as education, healthcare, and financial services [32, 33]. It's a structured yet flexible process that supports the creation of products that go beyond usability to deliver lasting value.

While User-Centered Design (UCD) emphasizes user needs and preferences, GDD takes a broader, goal-oriented perspective. It integrates user goals with design objectives, enabling designers to address deeper behavioral and contextual challenges [34]. This makes GDD a suitable choice for projects aiming to influence long-term user behavior and engagement.

### 2.4 Existing Financial Applications Evaluation

In today's saturated FinTech market, the launch of a new financial management application faces significant challenges due to intense competition [35]. Therefore, conducting a thorough competitor audit is essential to identify standard industry features, user expectations, and opportunities for differentiation. This analysis focuses on five widely used financial management applications: Snoop, Emma, Moneyhub, Spendee, and Pennies. These apps were selected based on interview data, popularity, user reviews, and the breadth of functionalities they offer.

The comparative analysis (Table 1) highlights common features such as expense tracking, budget management, bank integration, and dashboard visualisation. However, the findings also reveal gaps in areas like financial literacy support, gamification, and personalisation, which can be leveraged for innovation.

Snoop stands out for its spending insights and savings recommendations. Its strength lies in its recommendation engine, but it lacks features like investment tracking and customisation, limiting its appeal to advanced users.

Emma offers robust integration across bank accounts and financial services, positioning it as a versatile tool. However, it misses key engagement drivers such as financial education and tailored recommendations.

Moneyhub and Spendee provide strong financial analysis tools and investment tracking, yet neither addresses the educational dimension of financial management. This presents an opportunity for differentiation through embedded financial literacy content.

Pennies, in contrast, offers a highly intuitive and personalised budgeting experience. Despite this, the app lacks bank integration, a critical shortfall in a multi-account financial environment.

Overall, the analysis reveals that while existing apps offer solid functional foundations, few address user learning, motivation, or behavioral change. These insights support the development of a new solution that incorporates financial literacy and gamification to enhance user engagement and financial capability.

Table 1. The table of five competitor analysis.

| Functionalities | snoop | Emma | moneyhub | Spendee | pennies |
|---|---|---|---|---|---|
| Expenses tracking | ✓ | ✓ | ✓ | ✓ | ✓ |
| Budget management | ✓ | ✓ | ✓ | ✓ | ✓ |
| Financial literacy | ✓ | | | | |
| Bank integration | ✓ | ✓ | ✓ | ✓ | |
| Collaboration | | | | | |
| Dashboard | ✓ | ✓ | ✓ | ✓ | |
| Savings goals | ✓ | ✓ | ✓ | ✓ | |
| Bill tracking | ✓ | ✓ | ✓ | ✓ | |
| Subscription management | ✓ | ✓ | | ✓ | |
| Investment tracking | | | ✓ | | |
| Custom categories | | ✓ | ✓ | ✓ | ✓ |
| Notifications | ✓ | ✓ | ✓ | ✓ | |
| Multi-currency support | | ✓ | ✓ | ✓ | |
| Personalise | | | | | ✓ |
| Gamification | | | | | |
| Recommendations | ✓ | | ✓ | | |

## 3 Methodology

To address the research questions comprehensively, this study adopted a mixed-methods approach, combining both qualitative and quantitative techniques. Qualitative methods (e.g., interviews, think-aloud) were used to explore user motivations and behaviours in depth, while quantitative methods (e.g., surveys, SUS) provided generalisable insights and measurable outcomes.

### 3.1 Research Approach

**Semi-structured Interviews.** Used to explore user needs and financial behaviours through flexible, guided conversations. Eight international students at Edinburgh Napier University (aged 20–32) were interviewed to explore financial habits and literacy.

**Survey.** A 16-question online survey (n = 52) was distributed via social media to participants aged 18–30. It combined multiple-choice and open-ended items, designed to validate interview findings and explore financial behaviours at scale.

**Card sorting.** Closed card sorting was conducted remotely with 10 participants to inform intuitive information architecture, leveraging co-occurrence analysis to align with users' mental models (Fig. 1).

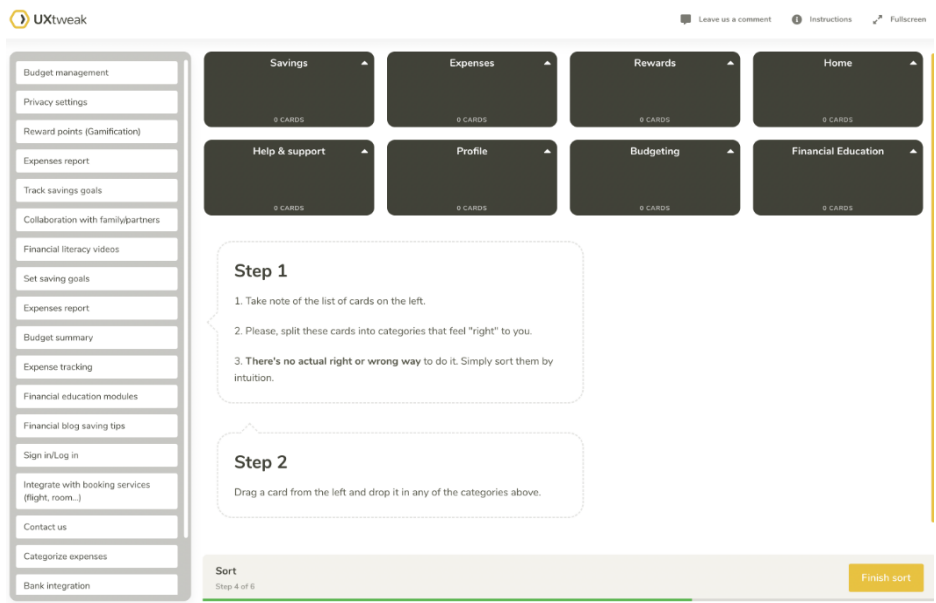

**Fig. 1.** Participant card sorting screen.

**Think Aloud.** The purpose of the think-aloud testing was to evaluate the usability and logic of the paper prototype by observing how users interact with it in a simulated environment. Six participants were provided with a list of tasks (Table 2) and guided to perform each while verbalising their thoughts, decisions, and emotional responses, allowing real-time observation of usability issues and cognitive friction. Sessions were recorded (without showing faces) and analysed for usability issues.

**Table 2.** Tasks assigned for participants in paper prototype testing.

| Number | Tasks |
|--------|-------|
| 1 | Please sign in to the app using username is user123, and password: user@123 |
| 2 | Integrate your account with Starling bank |
| 3 | Add a new expense of £45 Eating out by Cash |
| 4 | Set a saving goal for a Rome trip for £600, due October 2, 2024 |
| 5 | Go to the education section and choose to watch the video "How to save money?" |

**Usability Testing.** Conducted on high-fidelity prototypes to assess task completion, navigation, and pain points through observation and performance metrics. A different group of six students tested the high-fidelity prototype through six tasks (Fig. 2). Observations captured performance, confusion points, and feedback.

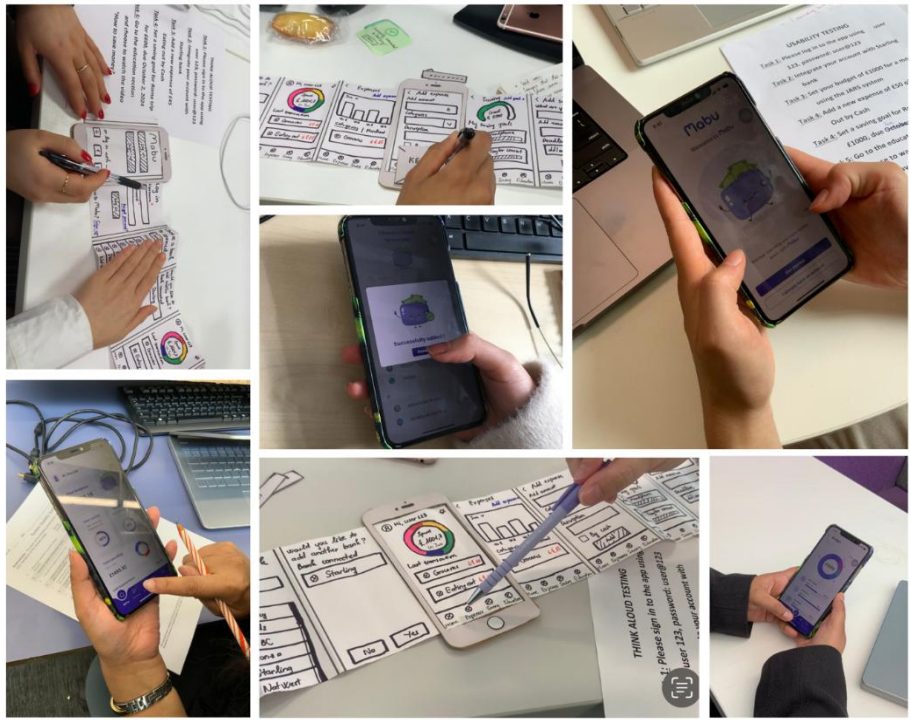

**Fig. 2.** User participation in the paper prototype and usability testing.

**System Usability Scale (SUS).** A post-test questionnaire provided a quantitative measure of perceived usability. Following usability testing, all six participants completed the SUS questionnaire. The results provided a standardised quantitative measure of perceived usability and were used to complement observational findings.

### 3.2 Ethical Consideration

All user research sessions strictly adhered to ethical standards. Participants were provided with a detailed consent form explaining the project's objectives, their role, and their right to withdraw at any point without consequence.

Informed consent was obtained prior to all interviews, surveys, and testing sessions. To ensure privacy, all responses were anonymised, and no identifiable personal data was collected. Sensitive financial questions were avoided to protect participants' comfort and well-being. Ethical transparency was maintained throughout the research process, ensuring participants understood how their input would be used exclusively for academic purposes within the scope of this study.

## 4 Design Of Mabu Application: Development of a Money Management App

### 4.1 User Research and Insights

User research combined eight semi-structured interviews and a survey with 52 respondents to explore financial behaviours among young people. Most participants tracked expenses via banking apps or spreadsheets, though some lacked any tracking method. Spending focused on lifestyle categories such as food, shopping, and transport. Many users struggled with saving due to limited financial knowledge and unstructured habits.

Survey results showed that users valued features like budget tracking, expense categorisation, and financial analytics. Gamification, notifications, and personalised recommendations were seen as motivators for continued engagement. Open-ended responses emphasised ease of use, visual clarity, and the need for automated tracking and bank integrations (Fig. 3).

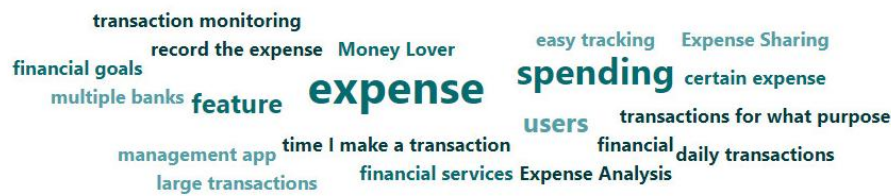

**Fig. 3.** Word cloud of features participants like to see in a financial app.

A closed card sorting study with 10 participants validated key app sections, including Budgeting, Savings, and Financial Education. Overall, findings informed both functional priorities and UX considerations, addressing the first two research questions regarding behaviours and challenges in financial self-management.

### 4.2 Designing Mabu - A Personal Financial App

**Persona.** Based on interview and survey findings, two personas were developed to represent the core user segments of the Mabu app: (1) a student with limited financial knowledge and (2) a young professional seeking more effective financial tools. This paper presents only one representative persona - Veronica (Fig. 4), a 28-year-old early-career professional with a stable income. She has basic budgeting habits but seeks smart features to track expenses, automate savings, and plan long-term goals. Veronica values usability, visual clarity, and personalised recommendations. Designing for her ensures the app meets the expectations of users who are financially conscious yet require digital support to optimise their financial planning.

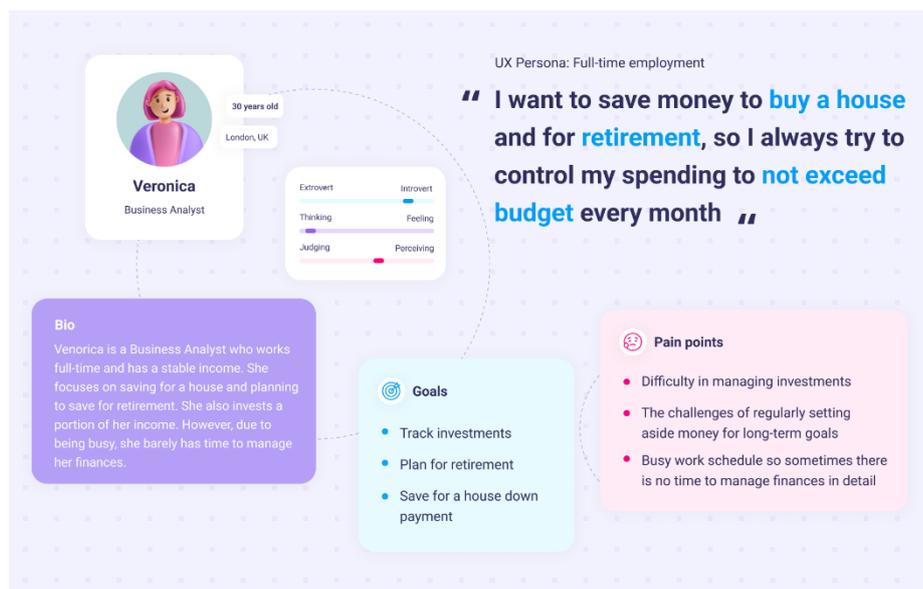

**Fig. 4.** Mabu's persona.

**Paper prototye.** Following the initial sketches and storyboard-based user journeys, a paper prototype of Mabu was developed with the aim of helping conduct early testing with potential users, uncovering potential errors early in the design process to improve and design high-fidelity prototypes, and helping to minimise changes later [36]. Key user flows - expense tracking, budgeting, and savings, were translated into hand-drawn interactive screens using paper, sticky notes, and cardboard to simulate a mobile interface. Elements such as buttons, text fields, and dropdowns were visually distinguished to guide user interaction (Fig. 5). This approach enabled rapid iteration and helped identify usability issues prior to developing the high-fidelity prototype.

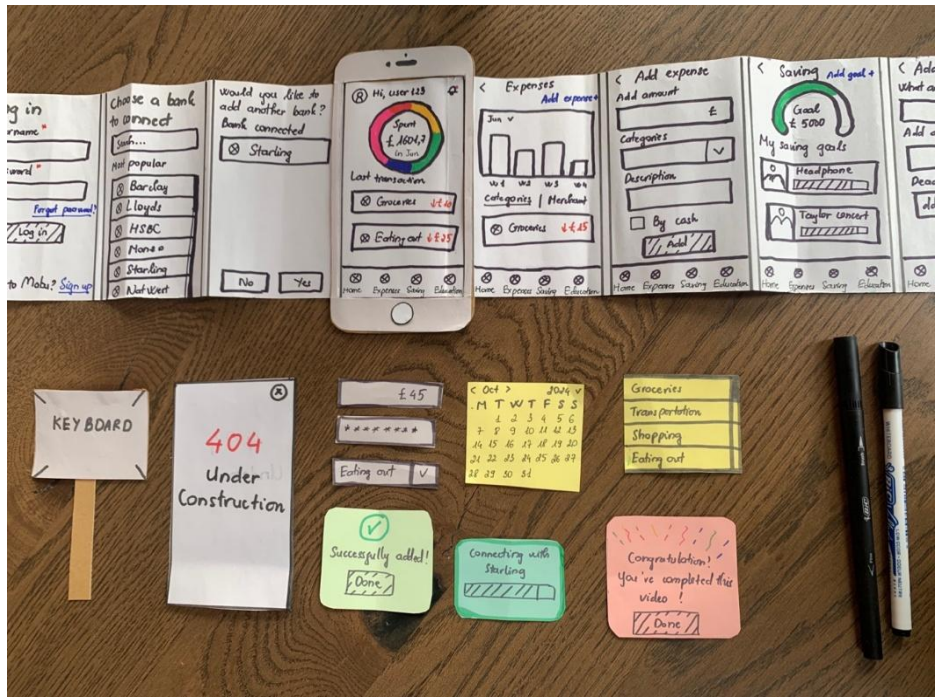

**Fig. 5.** Mabu's paper prototype.

**Hi-fi prototype.** The high-fidelity prototype of the Mabu application was developed in Figma, leveraging feedback from paper prototype testing to refine core user flows and UI elements. Key improvements focused on navigation clarity, visual hierarchy, and feature accessibility (Fig. 6).

To address usability issues identified in earlier testing, onboarding screens were redesigned: the "Login" and "Signup" labels were replaced with more intuitive phrasing. A savings countdown and budget progress tracker were added to the home screen to enhance user awareness and motivation. Expense screens now include trend charts and categorised spending breakdowns to support better financial decision-making.

Savings functionality was enhanced with progress bars and monthly breakdowns. The education section was enriched with videos, tips, and interactive quizzes to improve financial literacy. Login/signup screens follow Jakob's Law, aligning with familiar UI patterns, and a 6-digit verification step was added for security.

Bank integration features enable users to connect multiple accounts and view consolidated transactions, automatically categorised for ease of tracking. Budgeting functionality incorporates the JARS system, allowing users to allocate funds across six spending categories or customise their own.

To increase engagement, gamification elements such as badges were introduced to reward users for completing financial milestones (e.g., meeting savings goals or staying within budget). These features were added to encourage consistent interaction and reinforce positive financial behaviours.

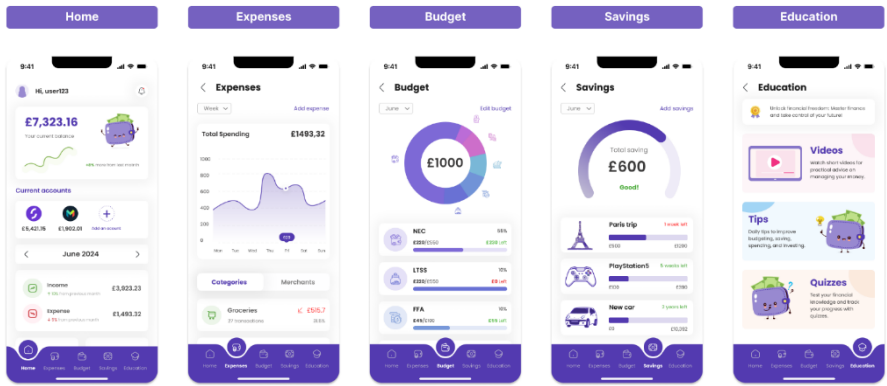

**Fig. 6.** Mabu's hi-fi prototype.

As part of an iterative design approach, additional improvements were implemented in response to both user feedback and heuristic evaluation. To support unfamiliar users, a three-step tutorial was added to explain the JARS budgeting system, along with navigation guidance for key app areas (Fig. 7).

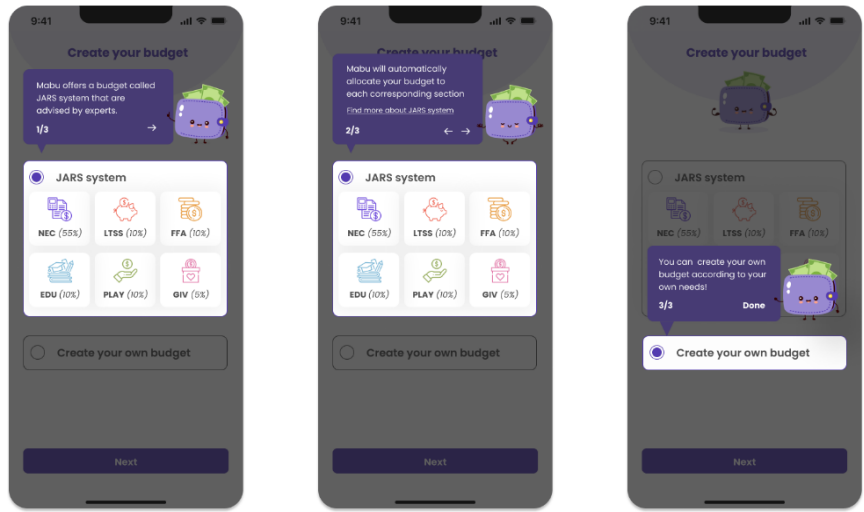

**Fig. 7.** Budget selection guide screens.

Expert feedback led to redesigning the bank connection flow, integrating popups for smoother user control, and improving error prevention via input validation and warning messages. Quick-access icons for creating new budgets, expenses, or savings were also added and pinned visibly across key interfaces to streamline frequent actions (Fig. 8). Together, these refinements demonstrate a continuous, user-centred design process that prioritises accessibility, usability, and motivation, laying a strong foundation for future evaluation and deployment.

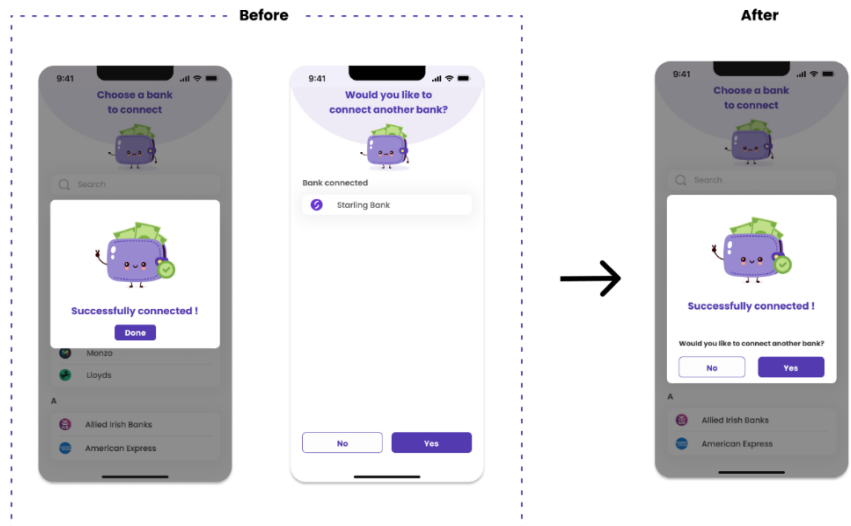

**Fig. 8.** Improved connecting to another bank screens.

### 4.3 Validation and User Feedback

Think-aloud testing with six participants revealed positive feedback on the app's usability, with most tasks completed successfully. However, issues emerged around task clarity and navigation. Users confused login and signup, struggled to locate expense and savings features, and recommended clearer buttons and a more intuitive layout. Participants also suggested adding an investment feature and visual progress tracking for savings goals.

Subsequent usability testing using the high-fidelity prototype confirmed similar issues. While login and bank integration tasks were completed quickly, challenges arose when setting budgets using the JARS system and adding new expenses. The "Add Expense" button was not prominent enough, and users experienced navigation friction. Task completion times further highlighted these issues, especially for expense tracking. Despite this, participants completed educational video tasks with ease.

The System Usability Scale (SUS), completed by all six participants, yielded an average score of 90, significantly above the industry benchmark of 68. This score indicates a high level of user satisfaction with the app's overall usability and design. Combined, these findings answered the final research question, showing that gamification elements - when balanced, enhanced engagement and motivation in managing personal finances. The SUS results of all six participants were summarised and calculated (Table 3) according to the following formula:

$$\text{SUS Score} = ((\Sigma\ (Q1, Q3, Q5, Q7, Q9)/5 - 5) + (25 - \Sigma\ (Q2, Q4, Q6, Q8, Q10)/5)) \times 2.5$$

In this case: Σ positive items (odd questions) = 129, Σ negative items (even questions) = 49

Adjusted score = (129/5 − 5) + (25 – 49/5) = 20.8 + 15.2 = 36

SUS Score = 36 × 2.5 = **90.0**

Table 3. Table of SUS from participants.

| Participant | Q1 | Q2 | Q3 | Q4 | Q5 | Q6 | Q7 | Q8 | Q9 | Q10 | Odd Q Total (x) | Even Q Total (y) |
|---|---|---|---|---|---|---|---|---|---|---|---|---|
| **P1** | 5 | 2 | 4 | 4 | 4 | 2 | 5 | 1 | 4 | 3 | 22 | 12 |
| **P2** | 4 | 2 | 4 | 2 | 5 | 2 | 5 | 1 | 4 | 1 | 22 | 8 |
| **P3** | 4 | 1 | 4 | 2 | 4 | 1 | 4 | 1 | 4 | 2 | 20 | 7 |
| **P4** | 5 | 3 | 4 | 2 | 2 | 1 | 4 | 1 | 3 | 4 | 18 | 11 |
| **P5** | 4 | 1 | 5 | 1 | 5 | 1 | 5 | 1 | 4 | 1 | 23 | 5 |
| **P6** | 4 | 1 | 5 | 1 | 5 | 1 | 5 | 1 | 5 | 2 | 24 | 6 |
| **Total** | | | | | | | | | | | **129** | **49** |

## 5   Discussion

The study identified several key findings. First, user research through interviews and surveys confirmed that young people lack adequate financial knowledge and seek tools that are easy to use, engaging, and educational. Features such as personalisation, interactivity, and embedded financial education were shown to enhance motivation and sustained use, aligning with previous research [1, 16].

Second, usability testing demonstrated that Mabu's design improved both engagement and literacy. Participants praised its intuitive interface and smart mascot, which contributed to a positive user experience. These results reinforce the principles of the Technology Acceptance Model (TAM) [17], where perceived usefulness and ease of use are crucial.

Third, the integration of Goal-Directed Design and gamification proved effective. GDD ensured alignment with user motivations, while gamification enhanced hedonic motivation and enjoyment [19, 25], resulting in increased user satisfaction and retention.

However, several limitations must be noted. The participant pool lacked diversity, as most users were based in Edinburgh, UK, which may affect generalisability. Time constraints limited the depth of iterative design, and the UXtweak tool used for card sorting had functionality limitations. Finally, the competitive audit reviewed only five financial apps, potentially overlooking emerging fintech innovations.

## 6 Conclusion and Future Work

This study employed a combination of semi-structured interviews, surveys, card sorting, and usability testing to explore young adults' financial behaviours and inform the design of Mabu. While interviews effectively uncovered user pain points, integrating contextual observation could have further enhanced insight. Surveys provided valuable data, though future iterations should avoid directly asking about desired features, as users often express wants rather than underlying needs. Closed card sorting contributed to a more intuitive information architecture but required deeper interpretation to align with user expectations. Usability testing - particularly the Think-Aloud method using paper prototypes - encouraged open feedback by reducing user pressure. The System Usability Scale (SUS) offered a standardised measure of user satisfaction, but some results may have been skewed due to user politeness or misunderstanding.

Future work will aim to expand user research across more diverse demographic groups, introducing features such as investment tools and personalised recommendations. Ethical considerations will be further prioritised through enhanced privacy policies, biometric authentication, and data encryption. Accessibility improvements will also be addressed. Overall, the use of Goal-Directed Design and gamification in Mabu shows strong potential in supporting financial literacy, though its long-term success relies on iterative development, ethical rigour, and continued user engagement.